# Multidimensional Light Detection with Symmetry-engineered Heterojunctions

Yucai Lin[1], Yaoqiang Zhou[1], Ruijuan Tian[1], Faisal Ahmed[1], Andreas C. Liapis[1], Youqiang Huang[1], Yawei Dai[1], Yuwei Chen[2], Weiwei Cai[3], Zongyin Yang[4], Weida Hu[5], Tawfique Hasan[6], Luojun Du[7*], Zhipei Sun[1*]

[1]Department of Electronics and Nanoengineering, Aalto University, Espoo, Finland.

[2]Hangzhou Institute of Advanced Studies, University of Chinese Academy of Science, Hangzhou, China.

[3]Key Lab of Education Ministry for Power Machinery and Engineering, School of Mechanical Engineering, Shanghai Jiao Tong University, Shanghai, China.

[4]College of Optical Science and Engineering, Zhejiang University, Hangzhou, China.

[5]State Key Laboratory of Infrared Physics, Shanghai Institute of Technical Physics, Chinese Academy of Sciences, Shanghai, China.

[6]Cambridge Graphene Centre, University of Cambridge, Cambridge, UK.

[7]Institute of Physics, Chinese Academy of Sciences, Beijing, China.

[*]zhipei.sun@aalto.fi, luojun.du@iphy.ac.cn

## Abstract

Miniaturised multidimensional light detection, encompassing full-Stokes polarimetry and spectroscopy in ultracompact footprints, is attracting growing interest for its potential to capture a comprehensive set of light's properties in portable platforms. Although significant progress has been made in miniaturised schemes for independent polarisation and spectral detection, achieving simultaneous high-dimensional light detection remains an outstanding challenge that limits their development toward full integration. We overcome this limitation by breaking the rotational and inversion symmetries in a symmetry-engineered van der Waals heterojunction to realise an ultracompact multidimensional photodetector. The dual symmetry breaking gives rise to non-trivial quantum geometric and topological features, enabling simultaneous broadband polarisation- and spectrum-resolved light detection, in contrast to previous van der Waals material-based devices, which could detect only one of these modalities. Our device, with an effective area of only ~10μm × 10μm, reconstructs full-Stokes polarisation with overall root-mean-square errors below ~0.05 and resolves spectral peaks separated by ~0.4 nm, capabilities not previously achieved in single-pixel detectors. By unifying high-fidelity polarimetry and sub-nanometre spectroscopy in a single electrically tunable junction, our work eliminates the need for cascaded detection architectures and establishes a foundation for multidimensional detector arrays for integrated photonics, quantum information processing, and precision imaging.

# Introduction

Multidimensional photodetectors that simultaneously capture information across multiple degrees of freedom, such as intensity, full-Stokes polarisation, and spectrum, are pivotal in enhancing our understanding and manipulation of light, catering to advancements across diverse applications such as coherent optical communication, material identification, precision agriculture, and quantum information processing[1–5]. As traditional photodetectors are typically limited to detecting one type of optical information at a time (e.g., intensity), the simultaneous acquisition of multidimensional light information requires multiple different detectors or complex setups involving cascaded optical elements. Therefore, multidimensional photodetection normally suffers from bulky footprints, alignment tolerances, and cumulative signal-to-noise penalties.

Recently, advances in computational spectral sensing have partially simplified the hardware architectures by reconstructing different light dimensions from a tunable pixel[6–16]. However, a fundamental limitation remains: the photoresponse tensor elements that would enable sensitivity to light's helicity or ellipticity are absent, as the active semiconductors are typically centrosymmetric and highly rotationally symmetric, making them insensitive to chiral or polarised light. For compact polarisation-state detection, prevailing strategies rely on patterning plasmonic metastructures atop photodetectors to induce the necessary symmetry breaking, integrating multiple complex detector modules[17–29], or employing anisotropic van der Waals (vdW) materials[11,30,31]. Yet, these approaches invariably increase fabrication complexity, introduce additional optical losses, restrict operational bandwidths, or lack intrinsic wavelength-resolving capability[11,30,31]. Consequently, truly miniaturised multidimensional light detection, despite recent progress[4,32] (see **Table S1**), remains elusive.

Here, we report a symmetry-engineered vdW heterojunction approach that achieves simultaneous broadband polarisation- and spectrum-resolved light detection, going beyond prior vdW material devices limited to either polarisation or spectral detection. By stacking two vdW crystals with different symmetries, we create an artificial heterojunction with the simultaneous breaking of rotational and inversion symmetries, and thus the emergence of non-trivial quantum geometric and topological properties, which are absent in the individual layers of the heterojunction. This, in conjunction with computational algorithms, enables the construction of multidimensional photodetectors that simultaneously capture the full Stokes polarisation states and spectral information over a broad wavelength range, owing to the symmetry-engineered photoresponse tensors that are fully polarisation-dependent and can be electrically tuned. Our results outperform current state-of-the-art multidimensional photodetectors by offering superior integration, scalability, and functionality, delivering a compact and high-efficiency solution well-suited for emerging and future photonic applications.

## Symmetry-engineering concept

Our concept stacks two crystals with a mismatched in-plane rotational order (**Fig. 1**a). The resulting heterojunction intrinsically exhibits broken inversion symmetry, due to the different constituent materials above and below the stacking interface[33]. Meanwhile, the rotational symmetry of the heterojunction is also reduced. Generally, stacking an $m$-fold crystal ($C_m$) on top of an $n$-fold one ($C_n$) reduces the interface symmetry to $C_x$, where $x$ is the greatest common divisor of $m$ and $n$. For example, the rotational symmetry of the formed heterojunction would be reduced to $C_1$ when materials with $C_2$ and $C_3$ symmetry are stacked together. The simultaneous breaking of inversion symmetry and reduction of rotational symmetry, in principle, can lead to the non-trivial quantum geometric and topological properties of electronic band structures (such as Berry curvature dipole), thereby enabling circular/linear photogalvanic effects (CPGE/LPGE), and allowing detection of both circularly and linearly polarised photons[10,34,35].

To discuss the above symmetry-engineering concepts in detail, we propose a vdW heterojunction example (**Fig. 1**b) by selecting γ-InSe (space group *R3m*, $C_3$ symmetry) and black phosphorus (BP, space group *Cmce*, $C_2$ symmetry) as two constituent materials[36,37]. We highlight that γ-InSe and BP are chosen based on the following considerations. First, the incompatible components and rotational symmetries between γ-InSe and BP break the inversion symmetry and reduce the rotational symmetry to $C_1$. It is noteworthy that although BP is an anisotropic material, the true linearly polarised light cannot be accurately determined. This is because of the existence of two-fold rotational symmetry and thus the indistinguishable photocurrents between incident polarisation angles $\theta$ and $(180°-\theta)$[11]. By contrast, the lowest rotational symmetry of $C_1$ can break this limitation, thereby detecting various linearly polarised incident light. Second, the photoresponse of a vdW heterojunction with broken inversion symmetry can be tuned by external electric fields, generating helicity- and ellipticity-sensitive photocurrent[38], and enabling the long-sought tunable photoresponses (CPGE/LPGE) required for full-Stokes polarimetry. Third, the γ-InSe/BP heterojunction exhibits a staggered-gap alignment at zero bias, which transitions to a broken-gap configuration under applied bias, enabling complex photoresponses with low noise and a distinctive spectral-density profile for precise spectral reconstruction[14,39–41]. Fourth, both γ-InSe and BP are direct bandgap semiconductors. The narrow bandgap (~0.3 eV) of BP and the wider bandgap (~1.4 eV) of InSe exhibit complementary optical and electronic characteristics, enabling photoresponses over a broad spectral range[8,40]. Therefore, symmetry-engineered vdW γ-InSe/BP heterojunctions offer extraordinary opportunities for multidimensional photodetectors to capture full-Stokes polarisation and spectral information simultaneously, resolving the long-standing degeneracy problem in highly symmetric photodetectors.

To experimentally realise our multidimensional photodetection concept, as shown in **Fig. 1**b, we (i) measure the electrically tunable photoresponses to light with known spectra and polarisation states, (ii) record the electrically dependent photoresponses to unknown light, and (iii) reconstruct the polarisation and spectral information of the unknown light using a trained neural network model, based on the reference responses obtained in Step (i). With this method, we integrate symmetry-engineered heterojunctions with neural-network decoding to simultaneously resolve full-Stokes polarisation and broadband spectrum without auxiliary metastructures, providing a compact platform for on-chip multidimensional sensing. Notably, this symmetry-engineering approach can also be extended to other material systems for advanced multidimensional light detection.

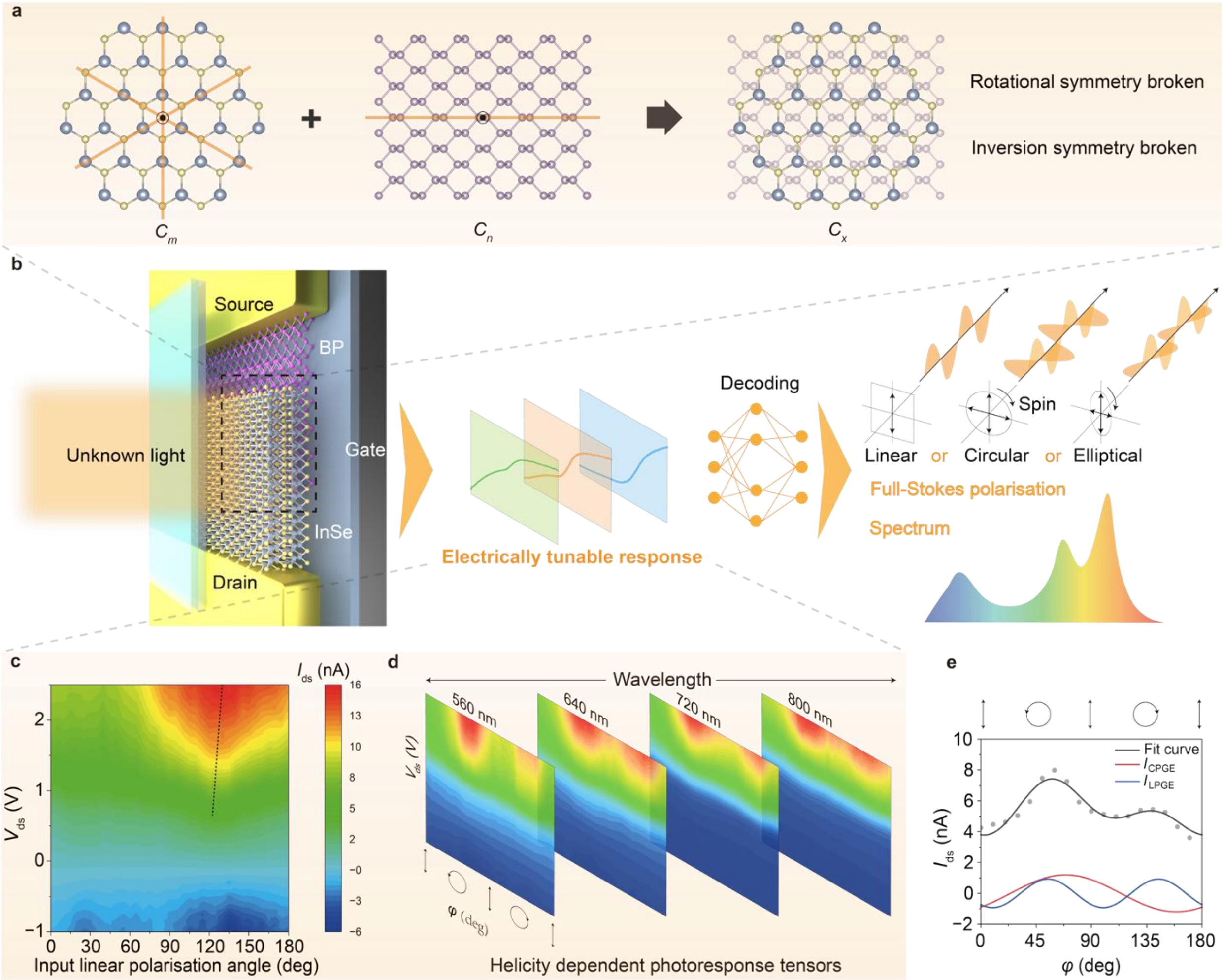

**Fig. 1 Symmetry-engineered heterojunctions for multidimensional light detection.** (a) Schematic illustration showing the stacking of two materials with symmetries $C_m$ (left) and $C_n$ (middle), resulting in the formation of a heterojunction with $C_x$ symmetry (right). Orange lines and circled dots represent mirror planes and rotational axes, respectively. (b) Schematic (Left) of a symmetry-engineered γ-InSe/BP heterojunction example. The right panel shows the electrically tunable photocurrent responses, which will be decoded to reveal multidimensional polarisation and spectral information simultaneously. (c) Photoresponse tensors for different incident linear polarisation angles ($\theta$) at the wavelength of 480 nm, showing that the linear polarisation yielding

the maximum photocurrent shifts with the applied bias voltage ($V_{ds}$) as indicated by a dashed line. (d) Photoresponse tensors measured under the light excitation with different wavelengths (560, 640, 720, 800 nm, left to right) and polarisation states following the sequence $\sigma^{o} \rightarrow \sigma^{-} \rightarrow \sigma^{o} \rightarrow \sigma^{+} \rightarrow \sigma^{o}$, and (e) the corresponding photocurrent at the wavelength of 540 nm, $V_{ds}$ = 2.5 V. The fitted components of circular photogalvanic effect (CPGE, red curve) and linear photogalvanic effect (LPGE, blue curve) from the generated bulk photogalvanic photocurrent (grey circles and curve). Grey circles denote experimental data, and solid curves are fitted results.

## Demonstration of Simultaneous Reconstruction of Full-Stokes Polarisation and Wavelength

Our γ-InSe/BP heterojunction device is fabricated from mechanically exfoliated γ-InSe and BP flakes inside a nitrogen-filled glovebox. The vdW heterojunction is characterised by Raman spectroscopy and atomic force microscopy (**Fig. S1**). The effective area of the heterojunction is ~10μm × 10μm. To improve the stability of γ-InSe and BP, the heterojunction is encapsulated with hexagonal boron nitride (*h*-BN) and $Al_2O_3$[42,43], as detailed in Methods. **Fig. 1c** presents the training photoresponse tensors recorded for various linear polarisation angles and bias voltages ($V_{ds}$) at the wavelength of 480 nm. The maximum photocurrent ($I_{ds}$) is observed at around 125°, as this orientation is parallel to the armchair direction of the BP flake. The maximum photocurrent shifts from ~124° to ~131° when $V_{ds}$ increases from ~1 V to 2.5 V. The resulting shifted and distinct photoresponse curves with different incident light polarisation states arise from the broken rotational symmetry at the γ-InSe/BP heterojunction. This aligns well with our previous discussion, with theoretical details provided in **Supplementary Note 2**. Thus, combining γ-InSe and BP into a symmetry-engineered heterojunction significantly enhances linear polarisation sensitivity while preserving efficient photoresponse characteristics[41,44].

For the chiral light detection test, part of the training photoresponse tensors for various incident chiral light inputs are shown in **Fig. 1**d. The helicity of the incident light is continuously modulated using a quarter-wave plate (QWP), as shown in **Fig. S2**. As $V_{ds}$ is swept, the device produces distinct, bias-controlled photocurrent signatures for each polarisation state across the measured spectral window, enabling unambiguous identification of linear, elliptical and circular components. **Fig. 1**e shows the photocurrent of the junction as a function of QWP angle at 540 nm. Over a 180° QWP modulation, the light sequentially transitions through linear ($\sigma^{o}$) → left circular ($\sigma^{+}$) → linear ($\sigma^{o}$) → right circular ($\sigma^{-}$) → linear ($\sigma^{o}$) polarisation states, and the photocurrent varies periodically with light helicity. The bulk photogalvanic photocurrent (grey circles and curve) could be extracted to helicity-dependent CPGE current (red curve) and the linear polarisation-sensitive LPGE current (blue curve). The observed CPGE photocurrents arise from a quantum-mechanical injection current, whereas the LPGE photocurrents originate from a quantum-mechanical shift current. Both phenomena reflect the geometric and topological electronic nature of the symmetry-engineered vdW heterojunctions(**Supplementary Note 3**)[10,34,35]. These results confirm that the electrically tunable responses of the symmetry-breaking heterojunctions to chiral polarisation, and allow possible differentiation of the incident light's polarisation states across a broad spectral range.

On the basis of polarisation-dependent training data, we use the above-mentioned reconstruction method to simultaneously measure the polarisations and wavelength of unknown incident light. **Fig. 2**a shows their reconstructed results for 12 different linear polarisation angles (left panel) and the corresponding wavelength accuracy (right panel) of incident light at ~480 nm. Note that the wavelength information is reconstructed simultaneously with the linear polarisation information. The root-mean-square error (RMSE) of the reconstructed 12 linear polarisation angles and wavelengths is ~3.33° and ~0.1 nm, respectively. The simultaneously reconstructed results at 560 nm and 620 nm are shown in **Figs. 2**b-c. The RMSE of the reconstructed linear polarisation angles and wavelength accuracy is ~4.44° and ~0.43 nm at 560 nm, and ~3.11° and ~5.52 nm at 640 nm, respectively. The results showcase the high fidelity for the reconstructed linear polarisation angle and wavelength, with more results for different wavelengths shown in **Fig. S5**.

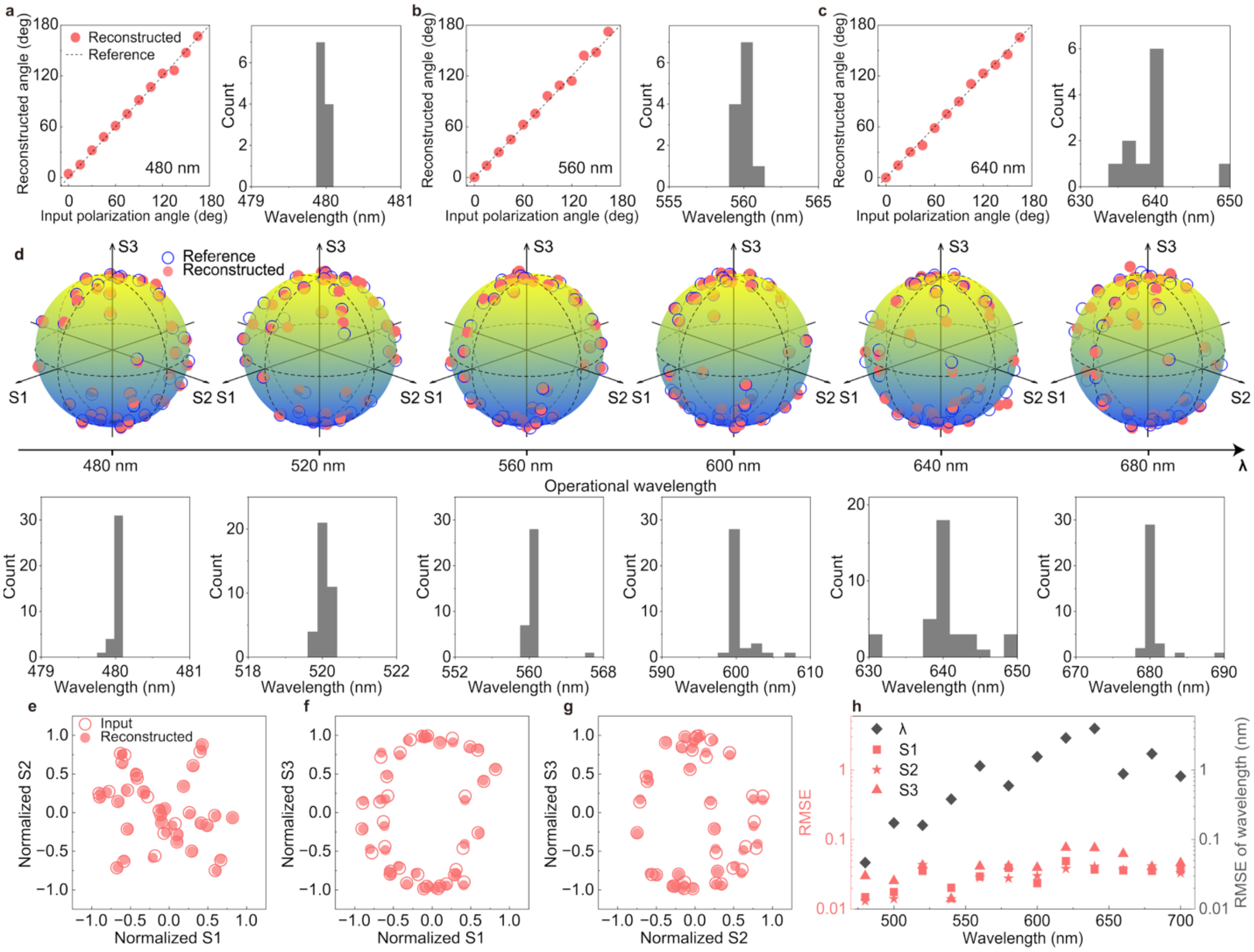

**Fig. 2 Simultaneous reconstruction of full-Stokes polarisation state and wavelength information.** Reconstructed linear polarisation angles (left) and corresponding reconstructed wavelength distributions (right) of incident light at the wavelength of (a) 480 nm, (b) 560 nm, and

(c) 640 nm. (d) Reconstructed full-Stokes polarisation states shown in the Poincaré sphere (upper) and corresponding reconstructed wavelength distributions (lower) for monochromatic incident light at different operational wavelengths ranging from 480 to 680nm, 36 different full-Stokes polarisation states for each wavelength. (e) S1-S2, (f) S1-S3, and (g) S2-S3 planes for reconstructed Stokes parameters for the incident light at 480 nm. (h) RMSEs of Stokes parameters (left) and wavelength (right) at different operational wavelengths.

To evaluate full-Stokes reconstruction, we test the device with circularly and elliptically polarised light spanning the complete Stokes space (S1, S2, S3). The reconstruction results of full Stokes polarisation and corresponding wavelength distributions for incident light at different wavelengths are shown in **Fig. 2**d. In the upper part, 36 different full-Stokes polarisation states for each wavelength are shown in the Poincaré sphere. Open circles represent the input reference data measured with a commercial polarimeter, and solid dots represent reconstructed values. The lower part presents their corresponding reconstructed wavelength accuracy of different polarisation states. Note that the full-Stokes polarisation and wavelength information are reconstructed simultaneously.

For 480 nm incident wavelength, the reconstructed Stokes parameters are projected onto the **Figs. 2** (e) S1-S2, (f)S1-S3, and (g) S2-S3 planes. The corresponding RMSEs for S1, S2 and S3 at 480 nm are ~ 0.015, 0.013 and 0.030, respectively. The overall RMSEs of S1, S2, S3, and wavelength for all 12 different wavelengths (36 different full-Stokes polarisation states per wavelength) are ~0.033, 0.034, and 0.051, respectively, as shown in **Fig. 2**h. The overall RMSE of the reconstructed wavelength accuracy is ~1.66 nm. The results show that the detector efficiently reconstructs full-Stokes parameters and the input wavelength simultaneously. Our demonstration achieves simultaneous retrieval of full-Stokes polarisation and wavelength information over the 480-700 nm range. More results for different wavelengths are shown in **Supplementary Note 4**. These full-Stokes parameter reconstruction errors are comparable to the state-of-the-art miniaturised full-Stokes polarimeters[4,10,21,26,45] (detailed comparison in **Table S2**). These results show that our ultracompact detector based on symmetry-engineered heterojunctions can accurately measure complex polarisation states, including circular and elliptical polarisation.

## Demonstration of High-resolution Spectral Reconstruction

Simultaneous reconstruction of the full-Stokes polarisation state and wavelength information enables comprehensive multidimensional light detection, although measuring a single optical parameter individually can yield superior performance in that specific dimension. For instance, here, we demonstrate the high-resolution spectral response of the photodetector by illuminating it with linearly polarised monochromatic and multi-peak light sources. Part of the wavelength-dependent photocurrent tensors used for training are shown in **Fig. S9**. After, we measure the photoresponses of unknown incident light and reconstruct their spectra using the reconstruction procedure described above (**Fig. 1**b). As shown in **Fig. 3**a and b, the reconstructed monochromatic spectra of unknown incident light in the 480-800 nm range measured with our symmetry-engineered photodetector match well with the reference spectra measured using a commercial spectrometer. The RMSE of the resolved peak wavelength accuracy is ~0.22 nm. Using the same method, we now measure unknown multi-peak spectra. The results are shown in **Fig. 3**c, with additional reconstructions of various complex multi-peak spectra provided in **Fig. S10**. All multi-peak spectra are reconstructed reliably from the electrical measurements. The device can also resolve dual-peak spectra with high accuracy. As shown in **Figs. 3**d and e, one peak is kept fixed at a wavelength of ~632.7 nm, while a second peak is tuned across the first one. Nevertheless, the ultracompact detector achieves a minimum resolution of ~0.4 nm across the measured spectrum, about an order of magnitude better than previously demonstrated miniaturised spectrometers[4,7,9,12,13,46] (see comparison in **Table S3**). We attribute this high performance to the staggered-gap alignment formed in the γ-InSe/BP heterojunction at zero bias, which shifts to a broken-gap configuration under applied bias. This transition enables complex photoresponses with low noise, supporting precise spectral reconstruction[14,39–41].

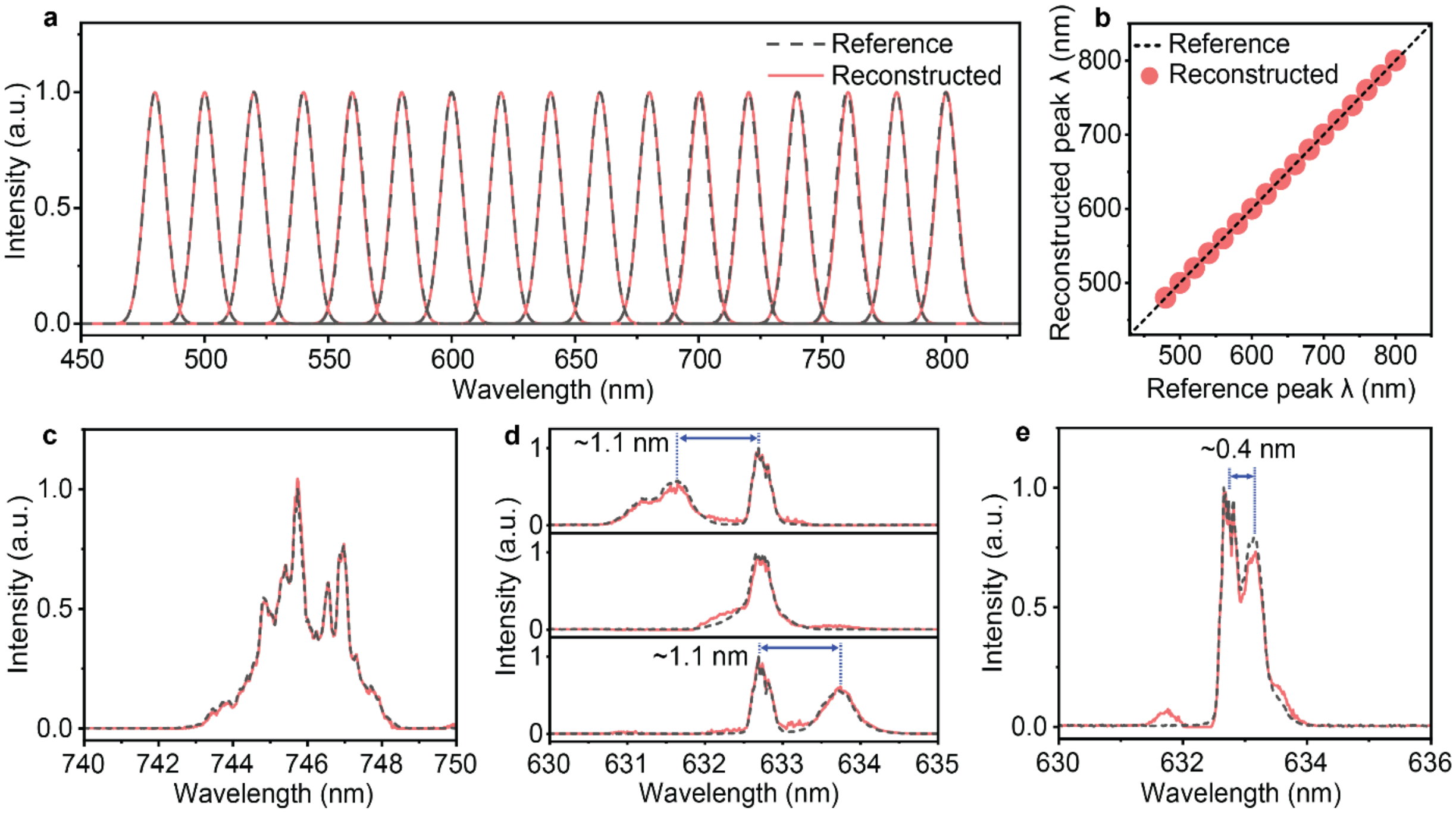

**Fig. 3 High-resolution spectrum reconstruction results.** (a) Monochromatic spectra and (b) the corresponding peak wavelengths (λ), and (c-e) multi-peak spectra reconstructed with our photodetector (solid curve) and input references measured with a commercial spectrometer (dashed curve).

These results show that our multidimensional photodetector combines exceptional polarisation and spectral sensitivity, covering the full-Stokes range and the entire visible spectrum. These demonstrations validate that the symmetry-engineered photodetector is promising to achieve the performance parameters typical of conventional detectors that measure either polarisation or spectrum individually. The ability to acquire multidimensional optical information in a single compact unit, without significant trade-offs in data quality, sets a new benchmark in intelligent photodetection[47]. This advancement significantly enhances our capacity to analyse and harness the multidimensional properties of light in scientific, industrial, and technological applications.

# Proof-of-concept multidimensional imaging

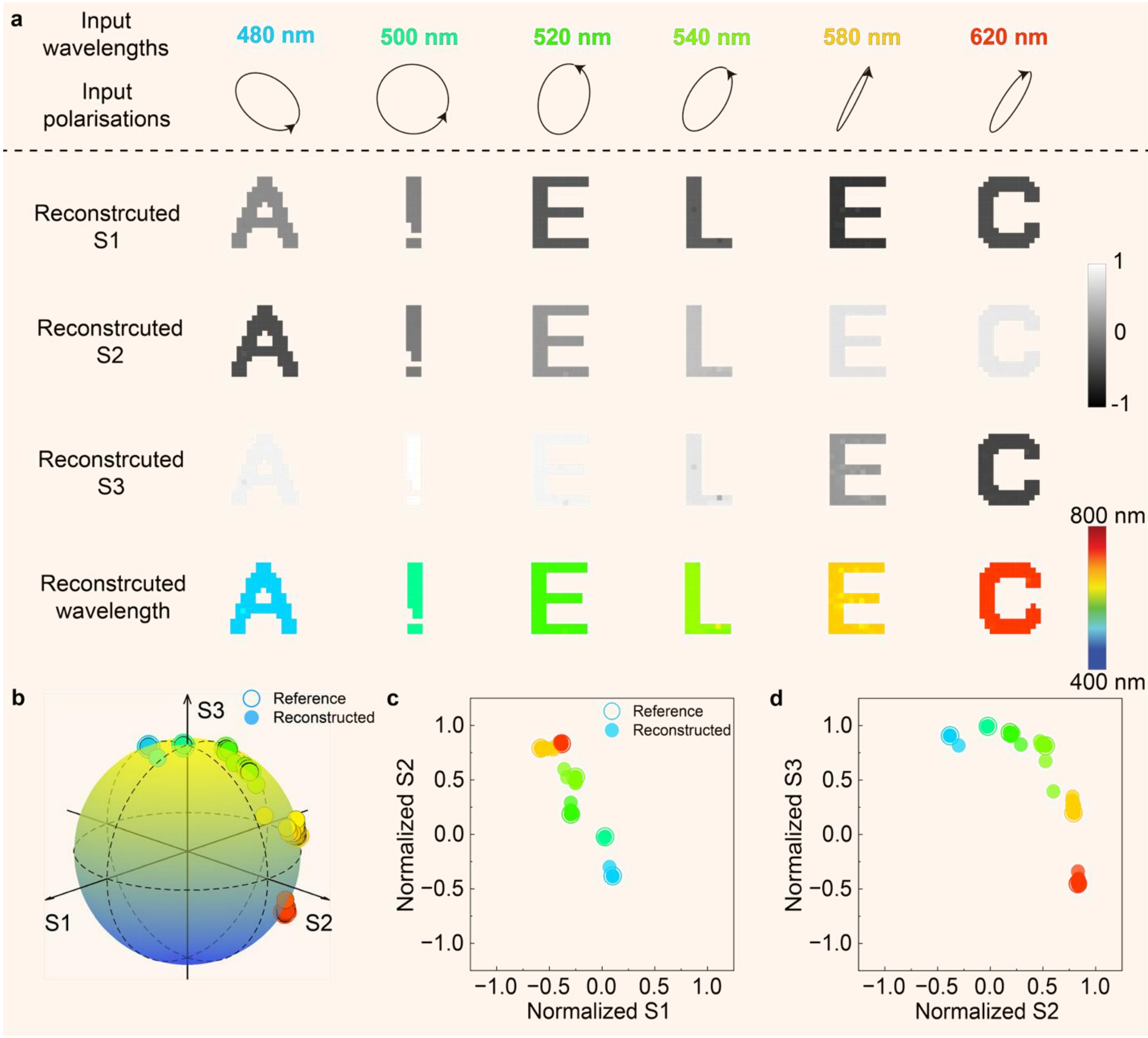

**Fig. 4 Proof-of-concept demonstration of multidimensional imaging.** (a) Multidimensional imaging of normalised Stokes parameters S1, S2, S3, together with the input wavelength. (b) All image pixels are shown with colour indicating wavelength and position on the Poincaré sphere indicating polarisation, and projected onto (c) S1-S2, and (d) S2-S3 planes. Each colored dot in (b-d) corresponds to the results at a specific incident wavelength (480, 500, 520, 540, 580, and 620 nm). The value of each dot represents the average of >37 measurements obtained at that wavelength across different pixels of the characters in (a).

Here, we perform proof-of-concept multidimensional imaging with our symmetry-engineered photodetector. We examine targets with variations in polarisation and spectral states, controlled by a linear polariser, half-wave plate, quarter-wave plate and tunable light source. The configuration of our measurement setup is shown in **Fig. S11**. **Fig. 4**a shows the reconstructed images as a false-colour mosaic, in which colour represents the wavelengths and grayscale represents the magnitude of the normalised Stokes parameters, all of which are reconstructed simultaneously. The crisp delineation of the letterforms, with each pixel preserving its spectral information while simultaneously rendering the complete polarisation states, highlights the detector's broadband, polarisation-sensitive responses. **Figs. 4**b-d show all the image pixels with colour indicating wavelength and polarisation state on the Poincaré sphere over 37 measurements per character. The RMSEs of S1, S2, S3, and wavelength accuracy for all the pixels are ~0.012, 0.009, 0.028, and 2.06 nm, respectively. All these variations are captured effectively by the photodetector. The present image sharpness is limited only by the scanning step; replacing the single-pixel demonstrator with a micro- or nanometre-pitch array would extend the same multidimensional capture scheme to megapixel-scale multidimensional imagers. Further, we evaluate the long-term reconstruction stability (see **Fig. S12**). The device shows wavelength reconstruction variation of ~3.57 nm within 31-day measurements, demonstrating reliable and robust device performance.

## Conclusion and Discussion

In summary, our results establish symmetry engineering as a powerful route for multidimensional light detection. By stacking γ-InSe ($C_3$) and BP ($C_2$) into a symmetry-mismatched heterojunction, we achieve concurrent breaking of rotational and inversion symmetries, generating a new class of photoresponse tensors with helicity and ellipticity sensitivity that are absent in the parent crystals. This mechanism transforms the ultracompact heterojunction into an electrically tunable encoder capable of resolving full-Stokes polarisation and sub-nm spectral features within a highly compact footprint. Combined with our neural-network-based reconstruction, the device unifies polarimetric and spectroscopic measurements without external optics or metastructures, greatly simplifying conventional architectures for multidimensional light measurements.

Beyond demonstrating a record level of multidimensional precision in a single-pixel detector, this work introduces symmetry as a controllable design axis for optoelectronic functionality, linking quantum geometric and topological effects to macroscopic photoresponse. The approach is general, scalable and compatible with state-of-the-art semiconductor and vdW platforms. Looking ahead, extending symmetry-engineered heterojunctions into electrically reconfigurable arrays and integrating them with adaptive learning models could enable intelligent photonic systems capable of interpreting light fields directly on chip. The ability to co-design physical symmetry, electronic response and computational inference opens new opportunities toward truly intelligent light-field sensors for secure communication and quantum-state read-out.

# Methods

## Device fabrication and characterisation

Bulk flakes of γ-InSe, BP, and hBN (2D Semiconductors) are used for exfoliation. After exfoliation, the flakes are transferred and stacked with the dry transfer method [48]. The stamp is passivated with an exfoliated hBN flake. The stacked heterojunction is transferred onto a Si/$SiO_2$ substrate with lithographically patterned contacts of evaporated Ti/Au (3/50 nm). The dry transfer is performed on a transfer stage inside a nitrogen-filled glovebox (MBRAUN), keeping oxygen and water concentrations below 0.1 ppm. To further shield the samples from ambient exposure, a 20-nm-thick $Al_2O_3$ layer is deposited at 120 °C by atomic layer deposition (ALD; Beneq TFS-500). Surface morphology is characterised by atomic force microscopy (AFM; Dimension Icon, Bruker). Raman spectra are characterised by a micro-Raman system (WITec Alpha300R) at room temperature.

## Electrical and optoelectrical measurements

A supercontinuum laser source (SuperK Extreme, NKT Photonics) combined with a tunable spectral filter (SuperK Varia, NKT Photonics) is used as a light source for training and testing. The spectral filter is set between 480 and 800 nm. The polarisation state of the output light from the supercontinuum laser source is controlled by a linear polariser, a half-wave plate, and a quarter-wave plate in cascade (**Fig. S2**). The reference full-Stokes parameters are measured by a polarimeter (Thorlab PAX1000). After passing through a 50× objective, the incident light was normally incident on our symmetry-engineered heterojunction. The focused spot is about 10 μm in diameter, closely matching the sizes of the heterojunctions. The drain-source voltage is swept from −1 V to +2.5 V, and the gate is set to either 0 V or 30 V. The photocurrent of the heterojunction at different drain-source voltages is collected with source meters (Keithley 2400). In the multi-peak measurements, the input light with multiple spectral peaks is generated by filtering the broadband output of the supercontinuum laser source with a home-built grating system. The reference spectra are measured by an Andor Shamrock 750 spectrograph equipped with an electron multiplying charge-coupled device (EMCCD; Newton 970). All measurements are performed at room temperature.

## Neural network and training details

TabPFN[49] is used to reconstruct the Stokes parameters and monochromatic spectra, whereas a convolutional neural network (CNN) is employed to reconstruct multi-peak spectra, with more details in **Supplementary Note 5**. The input layer of the networks consists of the voltage-dependent photoreponse tensor measured at arbitrary combinations of polarisation and spectral states. The output layer provides the predicted full-Stokes parameters (S1, S2, S3) and spectra. The spectra are normalised to the range 0-1. The CNN comprise three successive convolutional layers, each followed by a rectified linear unit (ReLU) activation. A max-pooling operation reduces dimensionality, after which a flattening layer prepares the data for fully connected (dense) layers. Dropout regularisation is applied before the dense layers to mitigate overfitting. The final dense layer output values for the mean-squared-error (MSE) regression task.

**Acknowledgements:** We acknowledge the provision of facilities and technical support from the Otaniemi research infrastructure (OtaNano-Micronova Nanofabrication Centre and OtaNano-Nanomicroscopy Centre), and the computational resources provided by the Aalto Science-IT project. We thank X. Cui, F. Nigmatulin, M. Uddin, F. Ali, J. Camilo Arias, and B. Liu for valuable discussions.

**Funding:** We acknowledge the financial support of the Finnish Ministry of Education and Culture through the Quantum Doctoral Education Pilot Program (QDOC VN/3137/2024-OKM-4), the Research Council of Finland (352780, 352930, 353364, 360411, 359009, 365686, and 367808), the Research Council of Finland Flagship Programme (320167, 358877), the EU H2020-MSCA-RISE-872049 (IPN-Bio), the Jane and Aatos Erkko foundation and the Technology Industries of Finland centennial foundation (Future Makers 2022), and ERC (834742).

**Author contributions:** Z.S. conceived of the ideas during discussions with Y.L., F.A. and A.C.L. Y.L. designed the experiments and carried out the characterisations and measurements. Y.L. fabricated the devices. F.A., Y.D., and Y.H. helped with the electrical and optoelectrical measurements. Y.L. developed the reconstruction code. Y.L., Y.Z., L.D., and Z.S. analysed the data. Y.C., W.C., Z.Y., W.H., and T.H. commented on the experimental results and helped with the data analysis. Y.Z., R.T., and Y.H. helped with the graphic design. Y.L., L.D., T.H. and Z.S. wrote the manuscript, and Z.S. supervised the research. All authors participated in the scientific discussion extensively and contributed to the writing of the manuscript.